\magnification=\magstep1
\baselineskip=16pt
\hfuzz=6pt

$ $

\vskip 1in

\centerline{\bf Uncomputability and physical law}

\bigskip

\centerline{Seth Lloyd}

\centerline{Mechanical Engineering, Massachusetts Institute of Technology}

\vskip 1cm

Computers can do so much that it's easy to forget that they were invented
for what they could not do.  In his 1937 paper,  
``On Computable Numbers, with an 
Application to the {\it Entscheidungsproblem,}" Alan Turing defined the
notion of a universal digital computer (a Turing machine), which became
the conceptual basis for the contemporary electronic computer [1].  Turing's
goal was to show that there were tasks that even the most powerful
computing machine could not perform.  In particular, 
Turing showed that no Turing machine could solve the problem
of whether a given Turing machine would halt and give an output
when programmed with a given input.  In creating a computational
analogue to G\"odel's incompleteness theorem [2], Turing introduced the
concept of uncomputability.

Although Turing's machine was an abstract mathematical construct,
his ideas soon found implementation as physical devices [3].  At the
end of the 1930s, Konrad Zuse in Germany began building digital
computers, first mechanical (the Z-1), and then electrical (the Z-3),
and Claude Shannon's 1937 master's thesis showed how digital
computers could be constructed using electronic switching
circuits [4], a development that presaged the construction of
the British code-breaking electronic computer, the Colossus,
and the American Mark I.    Since computers are physical
systems, Turing's uncomputability results show that there
are well-formulated questions that can be asked about physical
systems whose answers are uncomputable.  

That is, there are questions that we can ask about the physical
behavior of the universe whose answers are not resolvable by
any finite computation.  In this chapter, I will investigate the 
question of how such questions permeate the fabric of physical
law.  I will show that answers to some of the most basic questions 
concerning physical law are in fact uncomputable.  In particular,
one of the primary driving forces of science in general and of
physics in particular is to create the most succinct formulation
of the natural laws.  However, straightforward results from
algorithmic information theory imply that   
the most concise formulation of the fundamental laws of nature
is in fact uncomputable.  

Many physical systems are capable of universal computation: indeed,
it is difficult to find an extended system with nonlinear interactions
that is {\it not} capable of universal computation given proper
initial conditions and inputs [5-8].  For such systems, there are
always questions that one can ask about their physical dynamics
whose answers are uncomputable.  One might hope that such
questions are sufficiently esoteric that they do not overlap
with the usual questions that physicists ask of their systems.
I will show that even the answers to such basic questions
are often uncomputable.  In particular, consider the commonly
asked question of whether a physical system has 
an energy gap -- so that its energy spectrum is discrete
in the vicinity of its ground state, or whether it is gapless
-- so that the spectrum is continuous.  If the system
is capable of universal computation, then the answer to
the question of whether it has an energy gap or not is
uncomputable.

The uncomputability of the answers to common and basic questions of physics 
might seem to be a serious hindrance to doing science.   To the contrary,
the underlying uncomputability of physical law simply
puts physicisists in the same position that mathematicians
have occupied for some time: many quantities of interest
can be computed, but not all.  For example, even though
the most concise formulation of the underlying laws of physics
is uncomputable, short and elegant formulations of physical
laws certainly exist.  Not knowing
in advance whether or not the quantity that one is trying
to compute is uncomputable reflects the shared experience
of all scientists: one never
knows when the path of one's research will become impassible.
The underlying uncomputability of physical law 
simply adds zest and danger to an already exciting quest.

\bigskip\noindent{\it The halting problem}

Start by reviewing the origins of uncomputability.  Like
G\"odel's incompleteness theorem [2], Turing's halting problem
has its origins in the paradoxes of self-reference.  
G\"odel's theorem can be thought of as a mathematization
of the Cretan liar paradox.  The sixth-century BC
Cretan philosopher Epimenides
is said to have declared that all Cretans are liars.  
(As Paul says of the Cretans in his letter to Titus 1:12:
`One of themselves, even a prophet of their own, said,
The Cretians are alway liars, evil beasts, slow bellies. 
This witness is true.')
More concisely, the fourth-century BC Miletan philosopher
Eubulides, stated the paradox as `A man says, ``What I
am saying is a lie."'  The paradox arises because
if the man is telling the truth, then he is lying,
while if he is lying, then he is telling the truth.  

Potential resolutions of this paradox had been discussed
for more than two millenia by the time that Kurt G\"odel constructed his
mathematical elaboration of the paradox in 1931.  G\"odel 
developed a method for assigning a natural number
or `G\"odel number' to each well-formed statement or
formula in a mathematical theory or formal language.  
He then constructed theories that contained statements
that were versions of the liar paradox (`this statement
can not be proved to be true') and used G\"odel numbering
to prove that such a theory must be either inconsistent (false
statements can be proved to be true) or incomplete:
there exist statements that are true but that can not
be proved to be true within the theory.  

G\"odel's original incompleteness theorems were based on formal language
theory.  Turing's contribution was to re-express these
theorems in terms of a mechanistic theory of
computation.  The Turing machine is an abstract computing machine
that is based on how a mathematician calculates.  Turing
noted that mathematicians think, write down formulae
on sheets of paper, return to earlier sheets to look
at and potentially change the formulae there, and
pick up new sheets of paper on which to write.  A Turing
machine consists of a `head,' a system with a finite
number of discrete states that is the analogue to the
mathematician, and a `tape,' a set of squares each of which
can either be blank or contain one of 
finite number of symbols, analogous to the mathematician's
sheets of paper.  The tape originally contains a finite
number of non-blank squares, the program, which
specify the computation to be performed.  The head is prepared
in a special `start' state and placed at
a specified square on the tape.  Then as a function of
the symbol on the square and of its own internal state,
the head can alter the symbol, change its own internal
state, and move to the square on the left or right.
The machine starts out, and like a mathematician, the head
reads symbols, changes them, moves on to different
squares/sheets of paper, writes and erases symbols there,
and moves on.  The computation halts when the head enters
a special `stop' state, at which point the tape contains
the output to the computation.  If the head never enters
the stop state, then computation never halts.

Turing's definition of an abstract computing machine is
both simple and subtle.  It is simple because the components
(head and squares of tape) are all discrete and finite,
and the computational dynamics are finitely specified.
In fact, a Turing machine whose head has only two states,
and whose squares have only three states can be universal
in the sense that it can simulate any other Turing machine.
The definition is subtle because the computation can potentially
continue for ever, so that the number 
of squares of tape covered with symbols can increase
without bound.  One of the simplest questions that one
can ask about a Turing machine is whether the machine,
given a particular input, will ever halt and give an output, 
or whether it will continue to compute for ever.  Turing
showed that no Turing machine can compute the answer to
this question for all Turing machines and for all inputs: 
the answer to the question of whether a
given Turing machine will halt on a given input is uncomputable.


Like the proof G\"odel's incompleteness theorems, Turing's proof
of the halting problem relies on the capacity of Turing
machines for self-reference.   Because of their finite
nature, one can construct a G\"odel numbering for the set
of Turing machines.  Similarly, each input for a Turing
machine can be mapped to a natural number.  
Turing showed that a Turing machine that computes
whether a generic Turing machine halts given a
particular input can not exist.
The self-referential part of the proof consists of
examining 
what happens when this putative Turing machine evaluates
what happens when a Turing machine is given its
own description as input.  If such a Turing machine
exists, then we can define another Turing machine
that halts only if it doesn't halt, and fails to halt
only if it halts, a computational analogue to the Cretan liar
paradox (Saint Paul: `damned slow-bellied Turing machines').

The only assumption that led to this self-contradiction
is the existence of a Turing machine that 
calculates whether Turing machines halt or not.
Accordingly, such a Turing machine does not exist:
the question of whether a Turing machine halts or
not given a particular input is uncomputable.
This doesn't mean that one can't compute some of the
time whether Turing machines halt or not: indeed,
you can just let a particular Turing machine run,
and see what happens.   Sometimes it will halt and sometimes
it won't.  The fact that it hasn't halted yet even after
a very long time makes it unlikely to halt, but it
doesn't mean that it won't halt someday.
To paraphrase Abraham Lincoln,
you can tell whether some Turing machines halt all the time,
and you can tell whether all Turing machines halt some of the time,
but you can't tell when all Turing machines halt all of the time.

\bigskip\noindent{\it The halting problem and the energy spectrum
of physical systems}

At first it might seem that the halting problem, because of its
abstract and paradoxical nature, might find little application in
the description of physical systems.   I will now show, to the contrary,
that the halting problem arises in the computation of basic features
of many physical systems.  The key point to keep in mind is that even
quite simple physical systems can be capable of universal digital 
computation.  For example, the Ising model is perhaps the simplest
model of a set of coupled spins, atoms, quantum dots, or general
two-level quantum systems (qubits).  Arbitrary logical circuits 
can effectively be written into the ground state of the inhomogeneous 
Ising model, showing that the Ising model is capable of universal computation
in the limit that the number of spins goes to infinity [7].   When time-varying
fields such as the electromagnetic field are added, the number
of systems capable of universal computation expands significantly:
almost any set of interacting quantum systems is capable of universal
computation when subject to a global time-varying field [8].  

The ubiquity of computing systems means that the answers to many
physical questions are in fact uncomputable.  One of the most
useful questions that one can ask of a physical system is what
are the possible values for the system's energy, i.e., what is
the system's spectrum.   A system's spectrum determines many if
not most of the features of the system's dynamics and thermodynamics.
In particular, an important question to ask is whether a system's
spectrum is discrete, consisting of well-separated values for the
energy, or continuous.  

For example, if
there is a gap between the lowest or `ground state' energy
and the next lowest or `first excited state' energy,
then the spectrum is discrete and the system is said
to possess an energy gap.  By contrast, if the spectrum
is continuous in the vicinity of the ground state, then
the system is said to be gapless.  Gapped and gapless systems
exhibit significantly different dynamic and thermodynamic behavior.
In a gapped system, for example, the fundamental excitations or
particles are massive, while in a gapless sytem they are massless.
In a gapped system, the entropy goes to zero as the temperature
goes to zero, while for a gapless system, the entropy remains finite,
leading to significantly different thermodynamic behavior in
the low temperature limit.

As will now be seen, if a physical system is capable of universal
computation, the answer to the question of whether a particular
part of its spectrum is discrete or continuous is uncomputable.
In 1986, the Richard Feynman exhibited a simple quantum system whose
whose dynamics encodes universal computation [9].  Feynman's system
consists of a set of two-level quantum systems (qubits), coupled
to a clock.  The computation is encoded as a sequence of interactions
between qubits.  Every time an interaction is performed, the
clock `ticks' or increments by one.  In a computation that halts,
the computer starts in the initial program state and then
explores a finite set of states.  In a computation that doesn't
halt, the clock keeps on ticking forever, and the computer explores 
an infinite set of states.  In 1992, I showed that this feature
of Feynman's quantum computer implied that halting programs correspond
to discrete sectors of the system's energy spectrum, while
non-halting programs correspond to continuous sectors of
the system's spectrum [10].  In particular, when the system has
been programmed to perform a particular computation, the question
of whether its spectrum has a gap or not is equivalent to the
question of whether the computer halts or not.  

More precisely, a non-halting computer whose clock keeps ticking
forever corresponds to a physical system that goes through a 
countably infinite
sequence of distinguishable states.   Such a system by necessity
has a continuum of energy eigenstates.  Qualitatively, the derivation
of this continuum comes because the energy eigenstates are
essentially the Fourier transform of the clock states.  But
the Fourier transform of a function over an infinite discrete
set (labels of different clock states explored by the computation), 
is a function over a continuous, bounded set
(energy eigenvalues).   So for a non-halting program, the spectrum
is continuous.   Similarly, a halting computer that only
explores a finite set of distinguishable states has a discrete
spectrum, because the Fourier transform of a function over
a finite, discrete set (labels of the clock states)
is also a function over a finite discrete set (energy eigenvalues).
So for a halting program, the spectrum is discrete.
The reader is referred to [10-11] for the details of the
derivation.

Although the derivation in [10] was given specifically for Feynman's
quantum computer, the generic nature of the argument implies
that the answer to the question of whether {\it any} quantum system 
capable of universal computation is gapped or gapless is generically 
uncomputable.  If the computation never halts, then the system
goes through a countably infinite sequence of computational states
and the spectrum is continuous.  If the computation halts then
the system goes through a finite sequence of computational states and the
spectrum is discrete.
But since many if not most infinite quantum systems that evolve
according to nonlinear interactions are capable of universal computation,
this implies that uncomputability is ubiquitous in physical law.

\bigskip\noindent{\it The theory of everything is uncomputable}

As just shown, relatively prosaic aspects of physical systems,
such as whether a system has an energy gap or not, are uncomputable.
As will now be seen, grander aspirations of physics are also
uncomputable.  In particular, one of the long term goals of 
elementary particle physics is to construct `the theory of
everything' -- the most concise unified theoretical description
of all the laws of physics, including the interactions
between elementary particles, and gravitational interactions.
But the most concise description of any set of physical laws
is in general uncomputable.

The reason stems from the theory of algorithmic information [12-14].
The algorithmic information content of a string of bits is
the length of the shortest computer program that can produce
that string as output.  In other words, the algorithmic information
content of a bit string is the most concise description of that
string that can be written in a particular computer language.
The idea of algorithmic information was
first defined by Solomonoff [12] in order to construct a computational
version of Ockham's razor, which urges us to find the most parsimonious
explanation for a given phenomenon. (William of Ockham: {\it Numquam ponenda
est pluralitas sine necessitate} -- plurality should never be
posited without necessity, and {\it Frustra fit per plura quod potest 
fieri per pauciora} -- it is futile to do with more things what
be done with fewer.)   Kolmogorov [13] and Chaitin [14] 
independently arrived at the concept of algorithmic information. 

Algorithmic information is an elegant concept which makes precise
the notion of the most concise description.   In aiming for a 
theory of everything, physicists are trying to find the most
concise description of the set of laws that govern our observed
universe.  The problem is that this most concise description
is uncomputable.   The uncomputability of algorithmic information
stems from Berry's paradox.  Like all the paradoxes discussed
here, Berry's paradox arises from the capacity for self reference.
   
The English language can be used to specify numbers, e.g.,
`the smallest natural number that can expressed as the sum
of two distinct prime numbers.'   
Any natural number can be defined in English, and amongst
all English specifications for a given number, there is some
specification that has the smallest number of words.
Berry's paradox can be expressed as the following
phrase: `the smallest natural number that requires more than
twelve words to specify.'  Does this phrase specify a number?
If it does, then that number can be specified in twelve
words, in contradiction to the statement that it cannot
be specified in twelve words or less.   

The theory of computation makes Berry's paradox precise.  The shortest
description of a number is given by the shortest program on a given
universal Turing machine that produces that number as output.   Suppose
that algorithmic information is computable.  Then there is some program
for the same Turing machine which, given a number as input, outputs
the length of the shortest program that produces that number.   Suppose
that this program that computes algorithmic information content has 
length $\ell$.
Now look at the algorithmic version of Berry's phrase: `The smallest 
number whose algorithmic information content is greater than $\ell$
plus a billion.'   Does this phrase specify a number?  If algorithmic
information is computable, then the answer is Yes.  A program can 
compute that number 
by going through all natural numbers in ascending order, and computing
their algorithmic information content.   When the program reaches the
smallest number whose algorithmic information content is greater
than $\ell$ plus a billion, it outputs this number and halts.
The length of this program is the length of subroutine that computes
algorithmic information content, i.e., $\ell$, plus the length of
the additional code needed to check whether the algorithmic information
content of a number is greater than $\ell$ plus a billion, and if not, increment
the number and check again.  But the length of this additional code
is far less than a billion symbols, and so the length of the program
to produce the number in the algorithmic version of Berry's phrase
is far less than $\ell$ plus a billion.  So the algorithmic information
content of the number in Berry's phase is also less than $\ell$ plus
a billion, in contradiction to the phrase itself.  Paradox!

As in the halting problem, the apparent paradox can be resolved
by the concept of uncomputability.  The only assumption
that went into the algorithmic version of Berry's argument was that
algorithmic information content is computable.  This assumption lead
to a contradiction.  By the principle of {\it reductio ad absurdum},
the only conclusion is that algorithmic information content is
uncomputable.  The shortest description of a natural number or bit
string can not in general be computed.

The uncomputability of shortest descriptions holds for any bit string,
including the hypothetical bit string that describes the physical theory
of everything.  Physical laws are mathematical specifications of the
behavior of physical system: they provide formulae that
tell how measurable quantities change over time.  Such quantities
include the position and momentum of a particle falling under the force
of gravity, the strength of the electromagnetic field in the vicinity
of a capacitor, the state of an electron in a hydrogen atom, etc.
Physical laws consist of equations that govern the behavior of
physical systems, and those equations in turn can be expressed
algorithmically.  

There is a subtlety here.  Many physical laws
describe the behavior of continuous quantities.  Algorithmic
representations of the equations governing continuous quantities
necessarily involve a discretization: for example, using a finite
number of bits to represent floating-point variables.   
The concept of computability can be extended to the
computability of continuous quantities as well [15].
It is important to verify that the number of bits needed to 
approximate continuous behavior to a desired degree of accuracy
remains finite.   For the known laws of physics, such discrete
approximations of continuous behavior seem to be adequate in 
the regimes to which those laws apply.   The places where
discrete approximation to continuous laws break down -- e.g.
the singularities in the center of black holes -- are also
the places where the laws themselves are thought to break down.

The uncomputability of algorithmic information content implies
that the most concise expression of Maxwell's equations or the
standard model for elementary particles is uncomputable.  
Since Maxwell's equations are already concise -- they are
frequently seen on the back of a tee shirt -- few scientists
are working on making them even more terse.  (Note, however,
that terser the expression of Maxwell's equations, the more
difficult they are to `unpack': in all fairness, to assess
the algorithmic information content of a physical law,
one should also count the length of the extra computer code
needed to calculate the predictions of Maxwell's equations.)
In cases where
the underlying physical law that characterizes some phenomena
is unknown, however, as is currently the case for high $T_C$
superconductivity, then uncomputability can be a thorny problem:
finding even one concise theory, let alone the most concise one,
could be uncomputable.

Uncomputability afflicts all sciences, not just physics.  A simplified
but useful picture of the goal of scientific research is that scientists
obtain large amounts of data about the world via observation and
experiment, and then try to find regularities and patterns in that data.
But a regularity or pattern is nothing more or less than a method
for {\it compressing} the data: if a particular pattern shows up
in many places in a data set, then we can create a compressed version
of the data by describing the pattern only once, and then specifying
the different places that the pattern shows up.   The most compressed
version of the data is in some sense the ultimate scientific description.
There is a sense in which the goal of all science is finding theories
that provide ever more concise descriptions of data.

\bigskip\noindent{\it Computational complexity and physical law}

What makes a good physical law?  Being concise and easily expressed
is only one criterion.  A second criterion is that the predictions
of the law are readily evaluated.  If a law is concisely expressed 
but its predictions can only be revealed by a computation that takes 
the age of the universe, then the law is not very useful.  
Phrased in the language of computational complexity, if a physical
law is expressed in terms of equations that predict the future
given a description of the past, ideally those predictions can
be obtained in time polynomial in the description of the past state.
For example, the laws of classical mechanics and field theory
are described in terms of ordinary and partial differential equations
that are readily encoded and evaluated on a digital computer.

There is no guarantee that easily evaluated laws exist for
all phenomena.  For example, no 
classical computer algorithm currently exists that can predict
the future behavior of a complex quantum system.  The obstacle
to efficient classical simulation of quantum systems are the 
counterintuitive aspects of quantum mechanics such as quantum
superposition and entanglement which 
evidently require exponential amounts of memory space to represent
on a classical computer.  It seems to be as hard for classical computers
to simulate quantum weirdness as it is for human beings to comprehend it.

Embryonic quantum computers exist, however, and are capable of 
simulating complex quantum systems in principle [16].  Indeed, the largest
scale quantum computations to date have been performed by specialized
quantum information processors that simulate different aspects of
quantum systems.  In [17] Cory simulated the propagation of spin waves
in a sample of $O(10^{18})$ fluorine spins.  More recently, the D-Wave
adiabatic quantum computer has found the ground state of Ising models
using 512 superconducting quantum bits [18].  If we expand the set of 
computational devices that we use to define computational complexity,
then the consequences of the known laws of physics can in principle be 
elaborated in polynomical time on classical and quantum computers.

If we restrict our attention to laws whose consequences can be 
evaluated in polynomial time, then the problem of finding concise
expressions of physical laws is no longer uncomputable. 
The obstacle to finding the shortest program that produces
a particular data string is essentially the halting problem.  
If we could tell that a program never halts, then we could find
the shortest program to produce the data by going through all
potential programs in ascending order, eliminating as we go along all
non-halting programs from the
set of programs that could potentially reproduce the data.
That is, a halting oracle
would allow us to find the shortest program in finite time.

In the case where we restrict our attention to programs
that halt in time bounded by some polynomial in the length
of the inputs, by contrast, if the program hasn't halted
and reproduced the data
by the prescribed time we eliminate it from consideration
and move on to the next potential program.
This construction shows that the problem of finding the
shortest easily evaluated program is in the computational
complexity class NP: one can check in polynomial time
whether a given program will reproduce the data.  
Indeed, the problem of finding the shortest program
to produce the data is NP-complete: it is a 
variant of the NP-complete problem
of finding if there is {\it any} program within a specified subset
of programs that gives a particular output.
If we restrict our attention to physical laws whose predictions
can be readily evaluated, then the problem of finding the most
concise law to explain a particular data set is no longer uncomputable,
it is merely NP complete.  Probably exponentially hard is
better than probably impossible, however.

Similarly, in the case of evaluating the energy gap of
a physical system capable of computation, if the system
runs for time $n$, then the gap can is no greater than
$1/n^2$.   Accordingly, the problem of finding out
whether the gap is smaller than some bound $\epsilon$
is also NP-hard.

\bigskip\noindent{\it Discussion}

This chapter reviewed how uncomputability impacts the laws
of physics.  Although uncomputability as in the halting problem
arises from seemingly esoteric logical paradoxes, I showed that
common and basic questions in physics have answers that are
uncomputable.  Many physical systems are capable of universal
computation: to solve the question of whether such a system has
discrete or continuous spectrum in a particular regime, or whether
it is gapped or gapless, requires one to solve the halting problem.
At a more general level one can think of the all scientific laws
as providing concise, easily unpacked descriptions of obserbational
and experimental data.  The problem of finding the most concise
description of a data set is uncomputable in general, and the problem
of finding the most concise description whose predictions are easily
evaluated is NP-complete.  Science is hard, and sometimes impossible.
But that doesn't mean we shouldn't do it.

\vfill
\noindent{\it Acknowledgements:} The author would like to thank
Scott Aaronson for helpful discussions.

\vfil\eject
\noindent{\it References}

\vskip 1cm

\bigskip\noindent [1] A.M. Turing 
{\it Proc. London
Math. Soc.} {\bf 2 42}  230–265, {\bf 2 43} 544-546 (1937). 

\bigskip\noindent [2] K. G\"odel
{\it Monat.  Math. Phys.} {\bf  38}, 173–198 (1931).

\bigskip\noindent [3] M. Cunningham, {\it The History of Computation,}
AtlantiSoft, New York, 1997.

\bigskip\noindent [4] C. Shannon, {\it  A Symbolic Analysis 
of Relay and Switching Circuits,}
MIT MS thesis. 1937.

\bigskip\noindent [5] S. Lloyd, 
{\it Phys. Rev. Lett.} {\bf 75}, 346-349, (1995).

\bigskip\noindent [6] D. Deutsch, A. Barenco, A. Ekert,
{\it Proc. R. Soc. Lond. A} {\bf 8}, 669-677 (1995).

\bigskip\noindent [7] F. Barahona, {\it J. Phys. A} {\bf 15}, 3241 (1982).

\bigskip\noindent [8] S. Lloyd, {\it Science} {\bf 261}, 1569-1571 (1993).

\bigskip\noindent [9] R.P. Feynman, {\it Found. Phys.} {\bf 16},
507-531 (1986).

\bigskip\noindent [10] S. Lloyd,
{\it Phys. Rev. Lett.} {\bf 71}, 943-946 (1993).

\bigskip\noindent [11] T. Cubitt {\it et al.}, in preparation.

\bigskip\noindent [12] R.J. Solomonoff,
{\it Inf. Cont.},  {\bf 7}, 224-254 (1964).

\bigskip\noindent [13] A.N. Kolmogorov, {\it Prob. Inf. Trans.}, {\bf 1}, 3-11
(1965).

\bigskip\noindent [14] G.J. Chaitin, {\it J. Ass. Comp. Mach.} {\bf 13},
547-569 (1966).

\bigskip\noindent [15] L. Blum, M. Shub, S. Smale, {\it Bull. Am.
Math. Soc.} {\bf 21}, 1-46 (1989).

\bigskip\noindent [16] S. Lloyd, {\it Science}
{\bf 273}, 1073-1078, 1996.

\bigskip\noindent [17] W. Zhang, D. Cory, {\it Phys. Rev. Lett.}
{\bf 80}, 1324-1347 (1998).

\bigskip\noindent [18] M.W. Johnson {\it et al.} {\it Nature} {\bf 473}
194-198 (2011).

\vfill\eject\end